# Interface quality in GaSb/AlSb short period superlattices

Md Nazmul Alam[1], Joseph R. Matson[2], Patrick Sohr[1], Joshua D. Caldwell[2], and Stephanie Law[1a)]

[1]Department of Materials Science and Engineering, University of Delaware, 201 DuPont Hall, 127 The Green, Newark, Delaware 19716
[2]Department of Mechanical Engineering, Vanderbilt University, Nashville, TN 37235

a) Electronic mail: slaw@udel.edu

Heterostructures including the members of the 6.1Å semiconductor family (AlSb, GaSb, and InAs) are used in infrared optoelectronic devices as well as a variety of other applications. Short-period superlattices of these materials are also of interest for creating composite materials with designer infrared dielectric functions. The conditions needed to create sharp InAs/GaSb and InAs/AlSb interfaces are well known, but the AlSb/GaSb interface is much less well-understood. In this article, we test a variety of interventions designed to improve interface sharpness in AlSb/GaSb short-period superlattices. These interventions include substrate temperature, III:Sb flux ratio, and the use of a bismuth surfactant. Superlattices are characterized by high-resolution x-ray diffraction and infrared spectroscopy. We find that AlSb/GaSb short-period superlattices have a wide growth window over which sharp interfaces can be obtained.

# I. INTRODUCTION

The 6.1Å semiconductor family comprises AlSb, GaSb, and InAs. These materials have similar lattice constants, include type I, type II, and type III band offsets, and are fully miscible. These desirable properties have led to many applications for this family, including as infrared optoelectronic devices, as transistors and other electronic devices, and as testbeds for studies of fundamental physics[1–4]. Many applications rely on creating superlattices (SLs) of these materials, as SLs enable interactions between the quantum-confined states across multiple layers, resulting in devices with unique behaviors. However, for SLs to function as designed, the interfaces between the different materials must be atomically sharp.

There has been significant effort dedicated to improving interface quality in InAs/GaSb and InAs/AlSb SLs but relatively little investigation of AlSb/GaSb SLs. The quality of the AlSb/GaSb interface is important since this interface arises in devices that use all members of the 6.1Å family such as quantum cascade lasers and infrared photodiodes. In addition, short period AlSb/GaSb SLs may be used to tune the phonon modes and thus the SL infrared optical properties through the excitation of confined and interface phonons[5–8]. These structures could enable the rational design of new infrared materials[9,10], and high-quality AlSb/GaSb interfaces are needed.

Interface sharpness in SLs has been extensively studied in the related AlAs/GaAs system[11,12,21,22,13–20]. Authors have found that growth temperature, presence or absence of dopants, and III:As flux ratio play a role in determining the sharpness of the interface. As noted above, the related AlSb/GaSb system has been significantly less studied. It is expected that the growth kinetics of this system will be different than in the arsenide-based system, as antimony is known to act as a surfactant. In addition, AlAs is lattice-matched to GaAs, while AlSb has a 0.65% lattice mismatch with GaSb. In this paper, we explore how the sharpness of the interface in a GaSb/AlSb SL changes as a function of substrate temperature, III:Sb ratio, and the presence or absence of a bismuth surfactant. Samples are characterized by high-resolution x-ray diffraction and Fourier transform infrared spectroscopy.

## II. EXPERIMENTAL

The short period SLs studied in this paper were grown using a Veeco GENxplor molecular beam epitaxy (MBE) system and consisted of 20 monolayers (ML) of AlSb/4ML GaSb repeated 70 times for a total thickness of 500nm, as shown in Figure 1. All the samples were grown on (001)- oriented undoped GaSb substrates purchased from Wafer Technology Ltd. with a 68nm GaSb buffer layer and 5nm GaSb capping layer. The oxide layer was thermally desorbed at $(555 \pm 2)$ °C. The temperature was monitored by band edge thermometry (BET). We were able to obtain acceptable temperature control, perhaps due to the high quality of the GaSb substrates. Throughout the oxide desorption, the antimony pressure was kept at approximately $2 \times 10^{-6}$ Torr. The growth rate of AlSb and GaSb varied from 0.73~0.75 ML/S and 0.21~0.24 ML/S respectively. The layer thicknesses were determined from reflection high-energy electron diffraction (RHEED) oscillations.

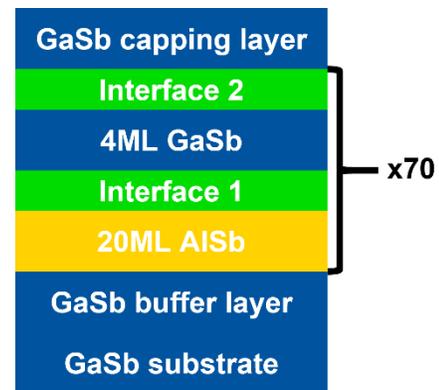

Figure 1. Sample schematic.

Three series of samples were grown to determine the optimal way to obtain sharp AlSb/GaSb interfaces. The first series of superlattices kept the Al:Sb and Ga:Sb flux ratio constant at 1:16 and 1:6 while varying the growth temperature $T_{growth}$=500˚C, 480˚C, and 460˚C. The second series of samples were grown at a constant growth temperature of 480˚C while the Ga:Sb and Al:Sb flux ratios (as determined by beam flux monitoring, BFM) were set to 1:2 and 1:9; 1:4 and 1:18; and 1:8 and 1:36, respectively. Finally, we investigated the effect of a bismuth surfactant by adding 20%, 10%, and 5% bismuth during the GaSb deposition to reduce the diffusion of gallium atoms into the adjacent AlSb layers[23–26]. In this case, the bismuth flux is reported as a percentage of the gallium flux. Bismuth is not expected to incorporate into the lattice under the growth conditions used here[27,28]. These samples were grown at a substrate temperature of 480˚C with Al:Sb and Ga:Sb flux ratios of 1:16 and 1:6, respectively. We also grew another sample in which the 10% bismuth flux was supplied during the deposition of both the GaSb and AlSb layers to compare the interface quality with the sample in which the 10% bismuth flux was applied only to the GaSb layers.

After growth, the superlattice quality was characterized by high resolution x-ray diffraction (HRXRD)[29]. The (004) reflection coupled scan was obtained using a Rigaku Ultima IV XRD system. The measured HRXRD curve was compared with simulated data to obtain the individual layer thicknesses, interface thickness, and composition. The modeling was performed using Globalfit software. In our modeling, the two-interface model has been used, where interface 1 and interface 2 are defined as GaSb-on-AlSb and AlSb-on-GaSb layer, respectively, as shown in Figure 1. The superlattices were further characterized by Fourier transform infrared (FTIR) spectroscopy. The FTIR reflection spectra were collected using a Bruker Hyperion 2000 microscope coupled to a Bruker Vertex 70v spectrometer. A closed-cycle, superconducting bolometer from QMC Instruments was used to gain access to the far-IR (<650cm$^{-1}$) with high sensitivity to clearly probe the optic phonons in the superlattices. The spectra were collected with a spectral resolution of 2cm$^{-1}$. See supplementary material at [URL will be inserted by AIP Publishing] for atomic force microscopy images of each sample. The sample surfaces were generally smooth, with root mean square roughness varying from 0.15nm to 0.56nm.

## III. RESULTS

The experimental HRXRD curves for all ten samples are provided in Figure 2 in gray. Figure 2(a), 2(b), and 2(c) show HRXRD curves of the samples grown with different substrate temperatures, different III:Sb flux ratios, and different amounts of bismuth surfactant respectively. We observe three classes of peaks: substrate peaks, the zeroth order SL peak, and satellite peaks. The substrate peak appears at 60.7°. An additional pair of small peaks appear at 54.3° and 53.7° and are instrumental artifacts. See supplementary material at [URL will be inserted by AIP Publishing] for an HRXRD scan of a bare GaSb substrate. The zeroth order SL peak appears to the left of the substrate peak near 60.3°. This shift is caused by the compressive strain of the superlattice with respect to the substrate due to the larger lattice constant of AlSb compared to GaSb (6.1355Å versus 6.09Å). The other peaks centered around the zeroth order peak are satellite peaks and are attributed to higher-order reflections from the SL. In Figures 2(a-c) these peaks are labeled with their reflection order. We observe satellite peaks up to +5 on the right side, but we only clearly observe the -1 satellite peak on the left side. The -3 peak appears in some scans but is often subsumed by the noise. This asymmetry is likely also caused by the compressive strain in the superlattice[30].

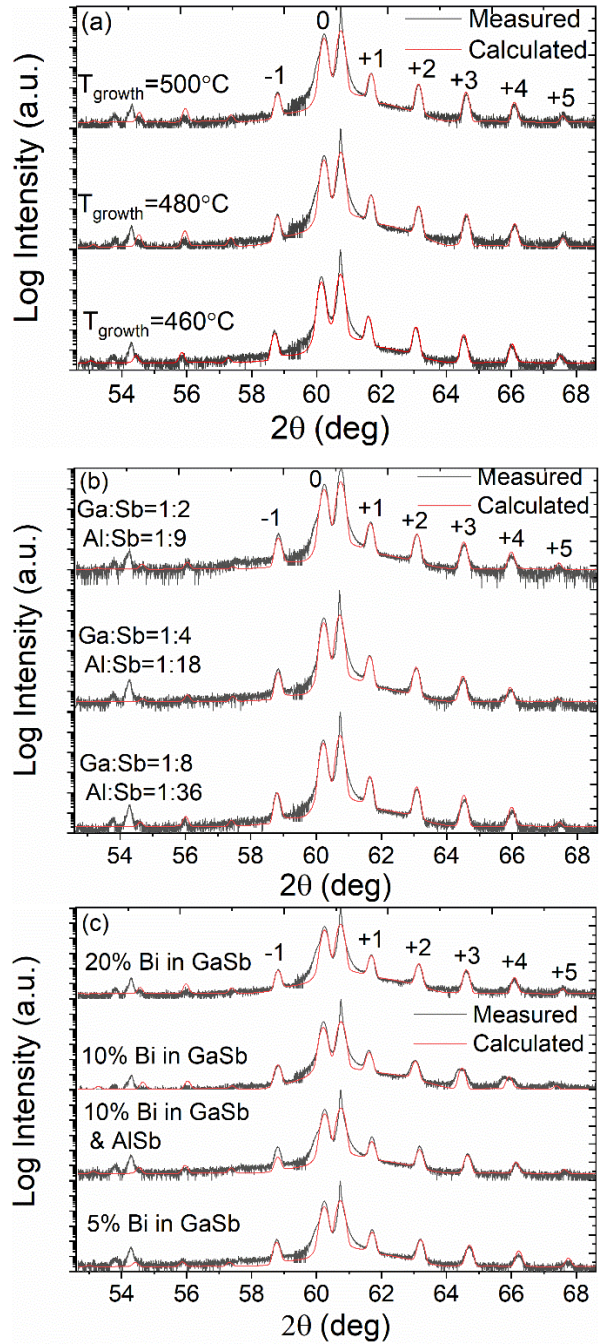

Figure 2 HRXRD (004) coupled scans for samples grown with (a) different substrate temperatures, (b) different III:Sb flux ratios, and (c) different amounts of bismuth surfactant. Numbers indicate diffraction peak orders. Gray lines are experimental data, red lines are modeling.

To extract the interfacial thickness and composition, we modeled the HRXRD data using the Globalfit software. When modeling the data, the layer thickness, layer composition, interface thickness, and interface composition were allowed to vary. Interface 1 and interface 2 were assumed to be different. The substrate miscut at 0.13°, intensity multiplier at 3.7, and strain at 0.39% were held constant in every simulation. Simulations indicated that the +5 and -4 order peaks were sensitive to the interface thickness and composition (see supplementary material at [URL will be inserted by AIP Publishing] for details on this peak sensitivity). Individual peak sensitivities to interfacial properties have been observed in other systems[31–33]. Since the -4 peak is weak and located at the same position as the machine artifact (near 54.5°), we focus on the +5 peak to determine the quality of the interfaces.

Figure 2 presents the fits for all ten samples shown in red. In all cases, the model fits the data well including the position, width, and intensity of the SL peaks. The substrate peak is narrower and more intense in the experimental data than in the model. This mismatch is attributed to a limitation in the software. The goodness of the fit was tested by modeling each experimental curve using different interface parameters while holding all other parameters constant. See supplementary material at [URL will be inserted by AIP Publishing] for examples. In these cases, we observed that all peaks except the +5 peak fit well. This demonstrates that the +5 peak is clearly sensitive to the interface quality and that our modeling is able to capture this sensitivity accurately.

In addition to HRXRD scans, we also performed FTIR spectroscopy measurements to probe the optical quality of the interfaces. The reflectance spectra are presented in Figure 3. Figure 3(a), 3(b), and 3(c) show the reflection spectra for the series of samples grown with different substrate temperatures, III:Sb flux ratios, and concentrations of bismuth surfactant, respectively. In all spectra, we observe the Reststrahlen bands of GaSb (229cm$^{-1}$ – 236 cm$^{-1}$) and AlSb (321 cm$^{-1}$ – 343 cm$^{-1}$), but with no additional spectral features, shifts, or changes in broadening that would indicate significant interfacial mixing [34,35].

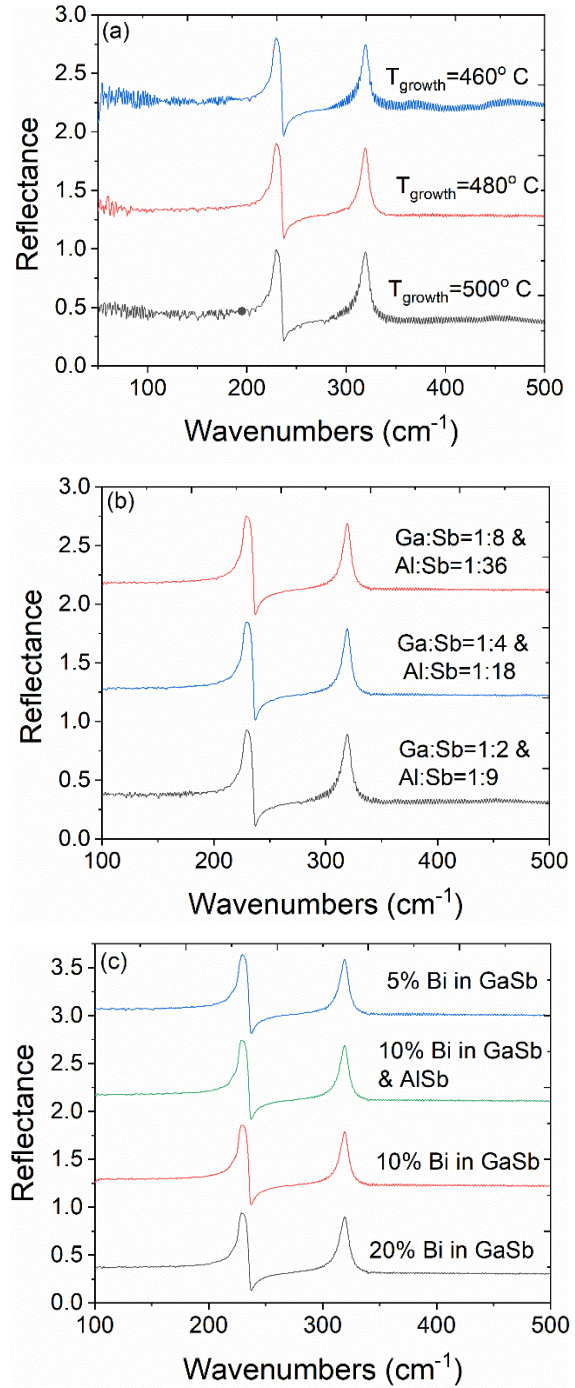

Figure 3. Reflection spectra for SLs grown with (a) varying substrate temperature, (b) varying III:Sb flux ratio, or (c) varying amounts of bismuth surfactant. Curves are offset for clarity.

## IV. DISCUSSION

The SL parameters extracted from the modeling of the HRXRD data are given in Table 1 including the AlSb layer thickness, the thickness and composition of Interface 1, the GaSb layer thickness, the thickness and composition of Interface 2, and the total thickness of one period found by adding the thicknesses of both layers and interfaces. The interface composition is described in terms of the percentage of aluminum in the Al$_x$Ga$_{1-x}$Sb alloy. We see that the AlSb and GaSb layer thicknesses are close to their intended values of 6.13nm and 1.22nm, respectively. However, the extracted layer thicknesses are smaller than the intended layer thicknesses for every sample. We expect one period of the SL to have a thickness of 7.35nm. If we instead compare the total thickness of one period to the intended period thickness, we see that these are quite similar. The small remaining deviation between the intended and actual period thicknesses is attributed to fluctuations in the MBE from day to day. This is supported by the fact that the three samples in each series were grown on a single day, and we observe similar period thicknesses for samples within a series.

Table 1 shows that the composition of Interface 1 is approximately evenly mixed, while Interface 2 is primarily composed of AlSb. This effect could be attributed to a difference in diffusion coefficients between the two materials and/or to asymmetric roughness. The diffusion coefficient for aluminum in GaSb is almost an order of magnitude larger than the diffusion constant for gallium in AlSb[36,37]. Asymmetric interface roughness may also play a role. Scanning tunneling microscopy measurements have shown that the AlSb-on-GaSb interface is rougher than the GaSb-on-AlSb interface[38]. A similar effect is observed in AlAs/GaAs interfaces, and it has been attributed to surface segregation of both gallium and of background impurities[39,40]. This mechanism may also exist in the GaSb/AlSb system. Interfaces that are dominated by AlSb also explains why the AlSb layer thickness deviates more from the intended thickness compared to the GaSb layer thickness: the "missing" AlSb can be found in the interfacial layers. We can also see that the interface thickness for all nine samples is always less than one unit cell and often less than one ML. When the interface is less than one ML, there are regions of the sample where the interface is atomically sharp and regions where it is alloyed, leading to an interface thickness that is nonzero, but less than one ML.

From the data presented in Table 1, we can conclude that none of the growth interventions had a meaningful impact on the sharpness of the interface. This is primarily because the interfaces were already quite sharp under normal growth conditions. This interpretation is supported by the FTIR data which shows that all nine samples have nominally identical phonon modes, all of which can be attributed to phonon modes in the pure GaSb and AlSb layers, with no strong evidence of phonon modes corresponding to the alloy. Overall, we can thus report that the growth widow for obtaining sharp interfaces in a GaSb/AlSb short period SL is quite large. Similar results have been found for GaAs/AlAs SLs[19,41]. However, it should be noted that in the GaAs/AlAs system that dopants have been found to significantly increase interface interdiffusion, leading to disordered interfaces[13–18]. It is possible that adding dopants to the GaSb/AlSb short period SL would also lead to similarly increased interfacial roughness, but that is beyond the scope of this study.

## V. SUMMARY AND CONCLUSIONS

We have grown AlSb/GaSb superlattices using molecular beam epitaxy under a variety of conditions in an attempt to improve the sharpness of the interfaces. Samples were characterized by high-resolution x-ray diffraction measurements and Fourier transform infrared spectroscopy measurements. HRXRD analysis indicated that all samples showed interfaces with relatively little roughness. This interpretation was supported by the identical FTIR reflection spectra observed for all samples. We conclude that the growth window for obtaining sharp interfaces in AlSb/GaSb SLs is quite large. This finding implies that growth of these SLs is compatible with growth of a wide range of other materials, enabling integration of antimony-based SLs with other semiconductor structures.

## ACKNOWLEDGMENTS

The authors acknowledge funding from the National Science Foundation, Division of Materials Research under Award No. 1904760 (Univ. Delaware) and Award No. 1904793 (Vanderbilt Univ.). M. N. A. and S. L. acknowledge the use of the Materials

Growth Facility (MGF) at the University of Delaware, which is partially supported by the National Science Foundation Major Research Instrumentation under Grant No. 1828141 and UD-CHARM, a National Science Foundation MRSEC under Award No. DMR-2011824.

## DATA AVAILABILITY

Data supporting the work presented is available from corresponding authors upon reasonable request.

## REFERENCES

[1] H. Kroemer, Phys. E **20**, 196 (2004).

[2] L.L. Chang and L. Esaki, Surf. Sci. **98**, 70 (1980).

[3] A. Rogalski, P. Martyniuk, and M. Kopytko, Appl. Phys. Rev. **4**, 31304 (2017).

[4] A.N. Baranov and R. Teissier, IEEE J. Sel. Top. Quantum Electron. **21**, 85 (2015).

[5] P. V Santos, A.K. Sood, M. Cardona, K. Ploog, Y. Ohmori, and H. Okamoto, Phys. Rev. B **37**, 6381 (1988).

[6] G.P. Schwartz, G.J. Gualtieri, W.A. Sunder, and L.A. Farrow, *Light Scattering from Quantum Confined and Interface Optical Vibrational Modes in Strained-Layer GaSb/AlSb Superlattices* (1987).

[7] M. Yano, T. Utatsu, Y. Iwai, and M. Inoue, J. Cryst. Growth **150**, 868 (1995).

[8] A. Milekhin, T. Werninghaus, D.R.T. Zahn, Y. Yanovskii, V. Preobrazhenskii, B. Semyagin, and A. Gutakovskii, Eur. Phys. J. B **6**, 295 (1998).

[9] J.D. Caldwell, I. Vurgaftman, J.G. Tischler, O.J. Glembocki, J.C. Owrutsky, and T.L. Reinecke, Nat. Nanotechnol. 2016 111 **11**, 9 (2016).

[10] D.C. Ratchford, C.J. Winta, I. Chatzakis, C.T. Ellis, N.C. Passler, J. Winterstein, P. Dev, I.

Razdolski, J.R. Matson, J.R. Nolen, J.G. Tischler, I. Vurgaftman, M.B. Katz, N. Nepal, M.T. Hardy, J.A. Hachtel, J.-C. Idrobo, T.L. Reinecke, A.J. Giles, D.S. Katzer, N.D. Bassim, R.M. Stroud, M. Wolf, A. Paarmann, and J.D. Caldwell, ACS Nano **13**, 6730 (2019).

[11] H. Fujimoto, C. Hamaguchi, T. Nakazawa, K. Taniguchi, K. Imanishi, H. Kato, and Y. Watanabe, Phys. Rev. B **41**, 7593 (1990).

[12] N. Sano, H. Kato, M. Nakayama, S. Chika, and H. Terauchi, Jpn. J. Appl. Phys. **23**, L640 (1984).

[13] W.D. Laidig, N. Holonyak, M.D. Camras, K. Hess, J.J. Coleman, P.D. Dapkus, and J. Bardeen, Appl. Phys. Lett. **38**, 776 (1981).

[14] J.W. Lee and W.D. Laidig, J. Electron. Mater. **13**, 147 (1984).

[15] S.W. Kirchoefer, N. Holonyak, J.J. Coleman, and P.D. Dapkus, J. Appl. Phys. **53**, 766 (1982).

[16] T.Y. Tan and U. Gösele, Appl. Phys. Lett. **52**, 1240 (1988).

[17] P. Mei, H.W. Yoon, T. Venkatesan, S.A. Schwarz, and J.P. Harbison, Appl. Phys. Lett. **50**, 1823 (1987).

[18] J.A. Van Vechten, J. Appl. Phys. **53**, 7082 (1982).

[19] R.M. Fleming, D.B. McWhan, A.C. Gossard, W. Wiegmann, and R.A. Logan, J. Appl. Phys. **51**, 357 (1980).

[20] N. Iwata, Y. Matsumoto, and T. Baba, Jpn. J. Appl. Phys. **24**, 17 (1985).

[21] J.A. Van Vechten, in *J. Vac. Sci. Technol. B Microelectron. Nanom. Struct.* (American Vacuum SocietyAVS, 1984), pp. 569–572.

[22] I. Lahiri, D.D. Nolte, J.C.P. Chang, J.M. Woodall, and M.R. Melloch, Appl. Phys. Lett. **67**, 1244 (1995).

[23] D. Kandel and E. Kaxiras, in edited by H. Ehrenreich and F. Spaepen (Academic Press, 2000), pp. 219–262.

[24] N. Grandjean, J. Massies, and V.H. Etgens, Phys. Rev. Lett. **69**, 796 (1992).

[25] J. Massies, N. Grandjean, and V.H. Etgens, Appl. Phys. Lett. **61**, 99 (1992).

[26] D. Wei, S. Maddox, P. Sohr, S. Bank, and S. Law, Opt. Mater. Express **10**, 302 (2020).

[27] Y. Zhong, P.B. Dongmo, J.P. Petropoulos, and J.M.O. Zide, Appl. Phys. Lett. **100**, 112110 (2012).

[28] G. Feng, K. Oe, and M. Yoshimoto, J. Cryst. Growth **301**–**302**, 121 (2007).

[29] P.F. Fewster, Acta Crystallogr. Sect. A Found. Crystallogr. **53**, 856 (1997).

[30] J.M. Vandenberg, J.C. Bean, R.A. Hamm, and R. Hull, Appl. Phys. Lett. **52**, 1152 (1988).

[31] Y. Zhou, J. Chen, Q. Xu, and L. He, J. Vac. Sci. Technol. B, Nanotechnol. Microelectron. Mater. Process. Meas. Phenom. **30**, 051203 (2012).

[32] D.M. Cornet, R.R. Lapierre, D. Comedi, and Y.A. Pusep, J. Appl. Phys **100**, 043518 (2006).

[33] D. Chrzan and P. Dutta, J. Appl. Phys. **59**, 1504 (1986).

[34] G. Lucovsky, K.Y. Cheng, and G.L. Pearson, Phys. Rev. B **12**, 4135 (1975).

[35] S.W. Da Silva, Y.A. Pusep, J.C. Galzerani, D.I. Lubyshev, A.G. Milekhin, V. V. Preobrazhenskii, M.A. Putiato, and B.R. Semjagin, J. Appl. Phys. **80**, 597 (1996).

[36] M. Gonzalez-Debs, J.G. Cederberg, R.M. Biefeld, and T.F. Kuech, J. Appl. Phys. **97**, 103522 (2005).

[37] J. Slotte, M. Gonzalez-Debs, T.F. Kuech, and J.G. Cederberg, J. Appl. Phys. **102**, 023511 (2007).

[38] J. Harper, M. Weimer, D. Zhang, C.-H. Lin, and S.S. Pei, Cit. J. Vac. Sci. Technol. B Microelectron. Nanom. Struct. Process. **16**, 1389 (1998).

[39] K. Leosson, J. Jensen, W. Langbein, and J. Hvam, Phys. Rev. B - Condens. Matter Mater. Phys. **61**, 10322 (2000).

[40] N. Chand and S.N.G. Chu, Appl. Phys. Lett. **57**, 1796 (1990).

[41] L.L. Chang and A. Koma, Appl. Phys. Lett. **29**, 138 (1976).

Table 1 The calculated individual layer thickness, interface thickness and composition $x$ in the $Al_xGa_{1-x}Sb$ alloy

| Sample | AlSb (nm) | Interface 1 | | GaSb (nm) | Interface 2 | | Total thickness (nm) |
|---|---|---|---|---|---|---|---|
| | | Thickness (nm) | Comp. (x) | | Thickness (nm) | Comp. (x) | |
| $T_{growth}$=500 | 5.45 | 0.20 | 0.44 | 1.13 | 0.29 | 0.89 | 7.07 |
| $T_{growth}$=480 | 5.49 | 0.11 | 0.71 | 1.18 | 0.28 | 0.81 | 7.06 |
| $T_{growth}$=460 | 5.55 | 0.13 | 0.88 | 1.13 | 0.25 | 0.74 | 7.06 |
| Ga:Sb=1:2 Al:Sb=1:9 | 5.49 | 0.37 | 0.43 | 0.98 | 0.39 | 0.85 | 7.23 |
| Ga:Sb=1:4 Al:Sb=1:18 | 5.52 | 0.51 | 0.51 | 0.99 | 0.24 | 0.56 | 7.26 |
| Ga:Sb=1:8 Al:Sb=1:36 | 5.55 | 0.20 | 0.48 | 1.25 | 0.19 | 0.82 | 7.19 |
| 20% Bi | 5.52 | 0.16 | 0.54 | 1.21 | 0.20 | 0.84 | 7.09 |
| 10% Bi | 5.38 | 0.36 | 0.62 | 1.00 | 0.49 | 0.71 | 7.23 |
| 5% Bi | 5.53 | 0.17 | 0.56 | 1.01 | 0.20 | 0.85 | 6.91 |

Figure 3: Sample schematic.

Figure 4: HRXRD (004) coupled scans for samples grown with (a) different substrate temperatures, (b) different III:Sb flux ratios, and (c) different amounts of bismuth surfactant. Numbers indicate diffraction peak orders. Gray lines are experimental data, red lines are modeling.

Figure 3. Reflection spectra for SLs grown with (a) varying substrate temperature, (b) varying III:Sb flux ratio, or (c) varying amounts of bismuth surfactant. Curves are offset for clarity.

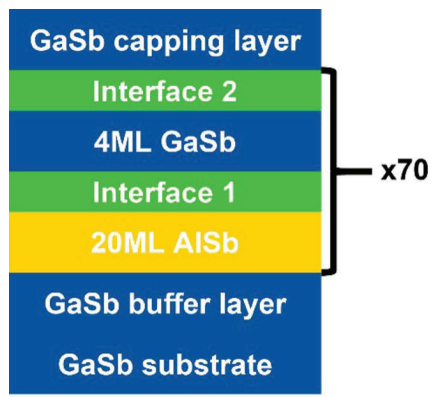

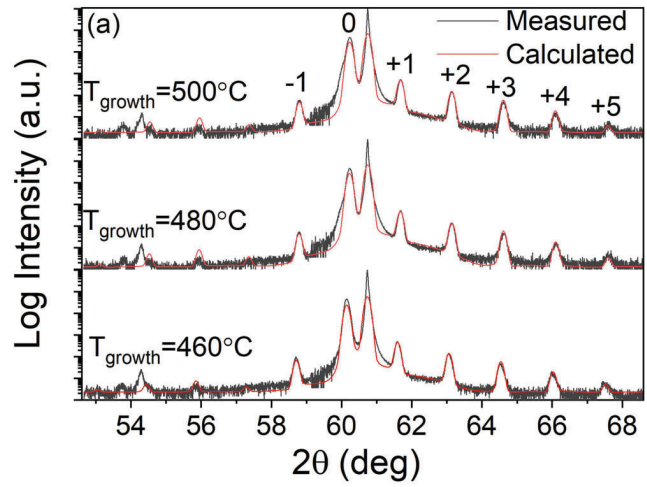
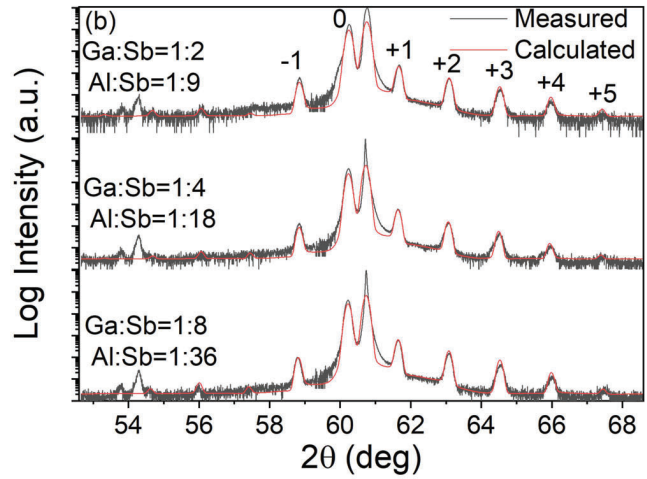
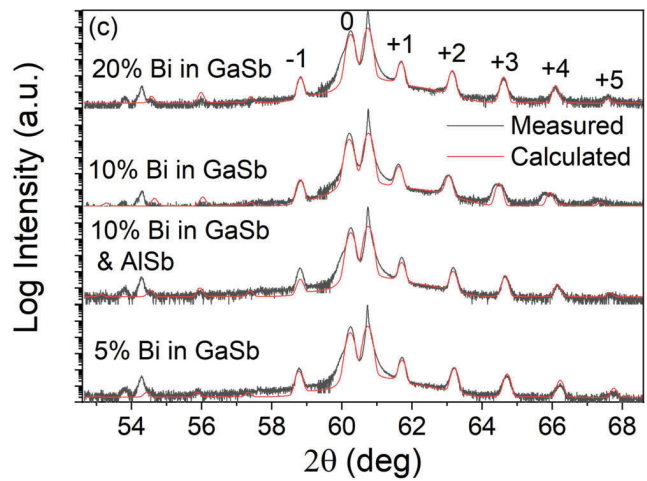

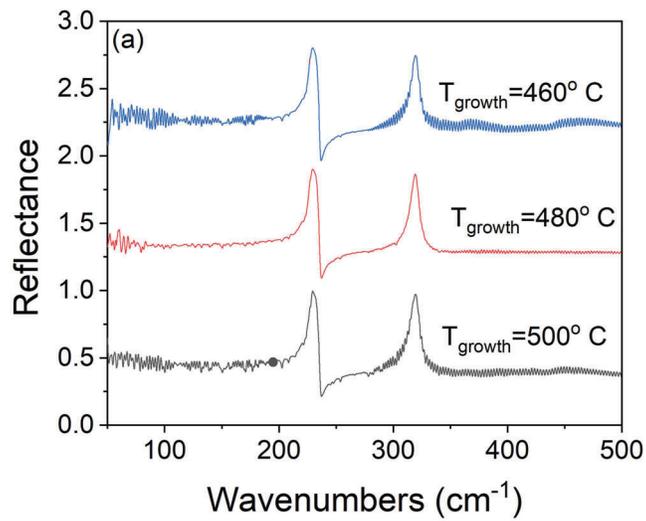
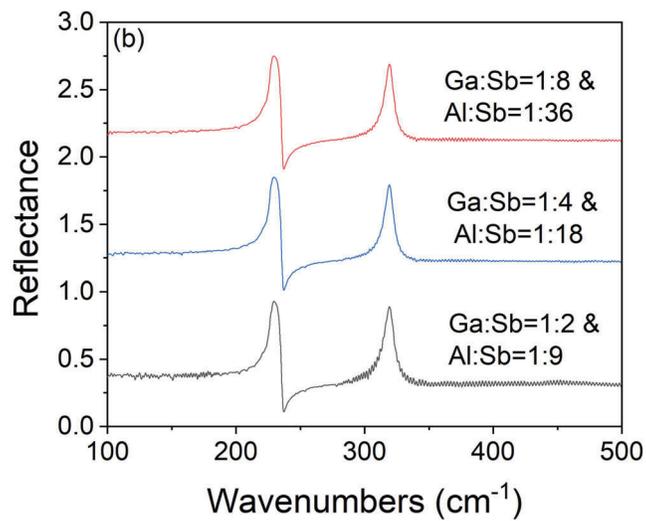
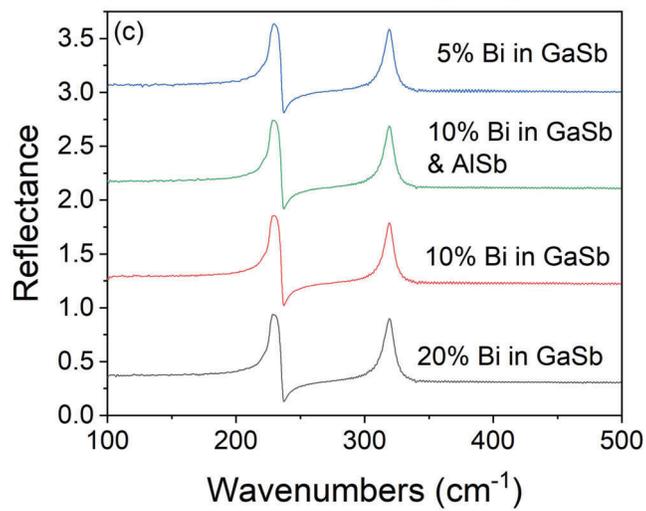